\documentclass[]{interact}

\usepackage[caption=false]{subfig}
\usepackage{lmodern}

\usepackage[longnamesfirst,sort]{natbib}
\bibpunct[, ]{(}{)}{;}{a}{,}{,}
\renewcommand\bibfont{\fontsize{10}{12}\selectfont}

\theoremstyle{plain}

\theoremstyle{definition}

\theoremstyle{remark}

\begin{document}


\title{\LARGE Filling the $uv$-gaps of the current VLBI   network in Africa}

 \author{
 \name{Marcellin Atemkeng\textsuperscript{a}\thanks{Corresponding author: Marcellin Atemkeng, Email: m.atemkeng@ru.ac.za}, Patrice M. Okouma\textsuperscript{b},  Eric Maina\textsuperscript{c}, Roger Ianjamasimanana\textsuperscript{c}, Serges Zambou\textsuperscript{d}}
 \affil{\textsuperscript{a}Department of Mathematics, Rhodes university, 6139 Grahamstown, South Africa.\\ \textsuperscript{b}College of Graduate Studies, University of South Africa, Johannesburg, South Africa.\\ \textsuperscript{c}Department of Physics and Electronics, Rhodes University, 6139 Grahamstown, South Africa.  \\ \textsuperscript{d}Department of Electrical and Electornics Technology, College of Technology, University of Buea, P.O. Box 63, Buea Cameroon}
 }
\maketitle
\begin{abstract}
In the African continent, South Africa has world-class astronomical facilities for advanced
radio astronomy research. With the advent of the Square Kilometre Array project in South Africa (SA SKA), 
six countries in Africa
(SA SKA partner countries) have joined South Africa to contribute towards the African Very
Long Baseline Interferometry (VLBI) Networks (AVN). Each of the AVN countries will soon have a single dish radio telescope that will be part of the AVN, the European VLBI Network, and the global VLBI network. 
The SKA and the AVN will enable very high sensitivity VLBI in the southern hemisphere.
In the current AVN network, there is a gap in coverage in the central African region. This work analyses the scientific impact if new antennas were to be built or old telecommunication 
facilities were to be converted to radio telescopes in each of the six countries in central Africa i.e. Cameroon, Gabon, Congo, 
Equatorial Guinea, Chad, Central African Republic. The work also discusses some economical and
skills transfer impacts of having a radio interferometer in this area of Africa.
\end{abstract}

\section{Introduction}
The spatial resolution of a particular telescope determines how well one can see details
of cosmic objects. This resolution depends on the size of the telescope and the wavelength
of the astronomical sources. Thus to have a better resolution, one solution would be to
build a telescope with a large diameter. However, there is a practical limitation on the
size of a telescope and this led to the development of interferometry. That is, instead of
having one large telescope, one can cross-correlate signals from individual antennas and
the resolution of the combined array of antennas (so-called interferometer) is determined
by the largest separation (baseline) between the individual antennas. The combined array
is therefore equivalent to a huge single-dish telescope with a diameter equal to the longest
baseline. Many radio telescopes were built after the development of interferometry technique.
However, by the mid-1960s \citep{Clark2003}, it was realised that some radio sources could not be
resolved even with radio telescopes of a few hundred km baselines. The quest for higher
resolution led to the development of the Very Long Baseline Interferometry (VLBI) \citep{Kellermann2001}. 
The VLBI is a technique of cross-correlating signals recorded by different antennas (and/or
array of antennas) separated by a large distance of up to the diameter of the earth. With
this technique, detailed images of astronomical objects at milliarcsecond resolution have
been obtained. In addition, high-precision astrometry has also been achieved.


The SKA \citep{Wild2017} will be split in a mid-frequency
(350 MHz-14GHz) part build in South Africa, which will incorperate MeerKAT \citep{Booth2012}, 
and a low frequency (50-350MHz) part in Australia. 
To enable high resolution interferometry through VLBI, the SKA South Africa currently leads an effort to 
convert existing unused telecommunication
dishes in partner countries (Botswana, Ghana, Kenya,
Madagascar, Mauritius, Mozambique, Namibia and Zambia) to radio telescopes. The converted antennas will then
become part of a network of antennas distributed throughout Africa to form the African
VLBI Network (AVN). Ghana has already successfully converted its old telecommunication
dish to a working radio telescope. Efforts to do the same in other African partner
countries are underway. The AVN will significantly improve the science capabilities of
the global VLBI community. The AVN combined with the existing international VLBI facilities
will produce huge quantities of data, presenting new challenges in data processing
and storage \citep{atemkeng2018baseline}. New techniques to manage the data must be developed, this includes:
storage systems and data compression techniques, machine learning methods, software
design techniques, control and monitoring systems that parallel the internet of things,
data flow architecture and systems dealing with massive scale computing. All of these
challenges will strengthen the scientific collaborations between South Africa and its partner
countries. In addition, Africa will become an international science and technology
focus.

The central African states are currently not part of the AVN. This paper investigates
the technical impact of this in terms of the AVN image quality and science capabilities.
We will demonstrate by means of simulations how the AVN image quality improves if
antennas were added in these countries. This paper also highlights the economical and 
technological benifits for these countries should they join the AVN project.

\vspace{1cm}

\section{Motivations}
The Central African States (ECCAS\footnote{Economic and Community of Central African States}) 
have not yet joined the AVN. However, a number
of opportunities will open up for these countries should they become part of the AVN. In
addition, as we will demonstrate in the next section, the existing AVN community will also
benefit a lot from the participation of these countries to the project. Below we highlights
potential benefits of joining the AVN.
\subsection{Education and Research impacts}
Joining the AVN will boost international cooperation in the field of Astronomy and engineering 
and enable participation in international scientific research.
Running a radio telescope requires skilled engineers, scientists and technicians who will manage
and run the facility. These personnels need to be trained in various disciplines, from
radio astronomy and astrophysics to computer science and engineering. The decision to
join the AVN network will trigger the development of critical skills and institutional capacity 
necessary to optimize ECCAS participation
in the SKA. High-end technologies and high performance computing facilities
needed to operate and maintain the telescopes are being developed in South Africa. The
Centre for High Performance Computing (CHPC) is already in place and running. 
\\\\
As mentioned before, Ghana has already successfully converted its old telecommunication
dish into a functioning radio telescope. This is a demonstration that Africa can participate
in high-level scientific research. The skills and experience from the development of
these facilities will be transferred to the ECCAS states if they join the AVN project. In
addition, the project will bring new scientific opportunities to the ECCAS countries on a
relatively short timescale. Currently, students in the AVN partner countries are benefiting
from a number of scholarships to pursue further studies, trained and acquired more skills.
Some of those scholarships includes but are not limited to: the SKA scholarship; 
the Development for Africa with Radio Astronomy (DARA) project in the
United Kingdom; a number of South African Research Chairs Initiative (SARChI)
of the Department of Science and Technology and the National Research Foundation.
Further benefit of joining the project includes the training of African scientists,
engineers and technicians, with a view to ensure that partner countries benefit from the
second phase of the SKA.
\subsection{Economic Impacts}
The AVN will trigger foreign investment and expenditure (including local contracts and suppliers). 
The skills development that will be promoted
by the project should enhance ECCAS engineering and scientific capabilities, promoting
science and engineering breakthroughs for other sectors such as medicine, remote
sensing and telecommunication, thereby enhancing innovation and competitiveness
among industries. The construction or conversion of the telescopes will pave the way
for a major boost to the local businesses e.g., tourism industries, construction industries,
hospitality industries, thus creating new job opportunities and enhancing local revenues.
\subsection{Scientific Impacts}
The  $uv$-coverage  of the current AVN combined with the European
VLBI Network (EVN)  have large unsampled $uv$-gaps (``holes'')  due to the absence of medium-length
baselines. The combined VLBI networks comprises two types of baselines: the
long baselines that link the EVN with the AVN antennas in South Africa and the short baselines from each of them.
The possible medium-length baselines from this combined instrument are from the single antenna in Ghana.
We need more medium-length baselines in between the longer and the shorter baselines to fill the ``holes'' in the $uv$-coverage. These
medium-length baselines can only come from linking the AVN and the EVN to antennas
from the central Africa or ECCAS countries.

\section{Radio interferometer, VLBI and $uv$-coverage}
The maximum resolution attenable by a single dish telescope with a diameter $D$, observing
at a wavelength $\lambda$ is approximately:
\begin{equation}
 \theta \sim \frac{\lambda}{D}. \label{eq1}
\end{equation}
Therefore, to achieve high-resolution observations, the diameter of the antenna must be
large or the observing wavelengths must be short. However, different scientific goals
require observations at different wavelengths. Thus, to resolve a radio source at a resolution
comparable with an optical telescope, astronomers use interferometry and aperture
synthesis techniques.
ki

In an interferometer the signals received by every antenna in the array
are cross-correlated witch each other, either in real-time or off-line,  these 
cross-correlation products are accumulated during a set period which is called 
the integration time. If the number of antennas is N then the instantaneous
number of correlation during the integration time is  $N(N-1)/2$. Note that the image 
quality depends on the number of cross-correlated signals. Because the relative orientation of the array and the sources 
change as the Earth rotates, one can take advantage of the Earth rotation to measure more samples \citep{Fomalont1973,Thompson1999}. 
This is called aperture synthesis, traditionally
known as the Earth rotation synthesis. The use of aperture synthesis techniques give a resolution comparable 
to that of a single dish telescope with a diameter equal to the longest
separation, $B_{max}$, between two antennas. Therefore, the $D$ in Eq. (\ref{eq1}) can now be replaced
by $B_{max}$. In order to achieve a milliarcsecond resolution  the network of
antennas requires baselines longer than $10^4$ km. Achieving such a high-resolution observation requires a VLBI technique.

Eq~(\ref{eq2}) shows  the differential of the spatial frequencies measured in wavelength  as a function of the baseline vector with  components $L_x$, $L_y$, $L_z$ along the axes of the ITRF coordinate system~\citep{bridle1999synthesis}. The baseline is tracking a  source at declination $\delta$ and hour angle $h$:
\begin{align}
\begin{split}
 \frac{\partial u}{\partial t}  &=\frac{\omega_e}{\lambda}\big( L_x \cos h - L_y\sin h\big)
\\
  \frac{\partial v}{\partial t} &=\frac{\omega_e}{\lambda}\big( L_x\sin\delta \sin h +L_y \sin\delta \cos h\big)\\
 \frac{\partial w}{\partial t} &=\frac{-\omega_e}{\lambda}\big( L_z\cos\delta \sin h +L_y \cos\delta \cos h\big) ,\label{eq2}
\end{split}
\end{align}
where $\omega_e = 7.2925 · 10^{−5}$ $rad.s^{-1}$ is the angular velocity of the Earth. The $uv$-coverage is the set of all the projected baseline vectors, $(u,v,w)$ in the Fourier plane or $uv$-plane. In other words, the $uv$-coverage describes the Fourier transform of the  interferometer response
toward a source at the phase centre of the observer.
An efficient way to fill the $uv$-coverage is to add many antenna telescopes together, while making
use of the Earth's rotation, the frequency coverage and antennas layout of the interferometer. The more complete the
$uv$-coverage, the better the response of the instrument. 

\subsection{Why the missing samples in the EVN + Kuntunse + MeerKAT  $uv$-coverage?}
\label{sect1}
\begin{figure}
  \centering
    \includegraphics[width=.6\textwidth]{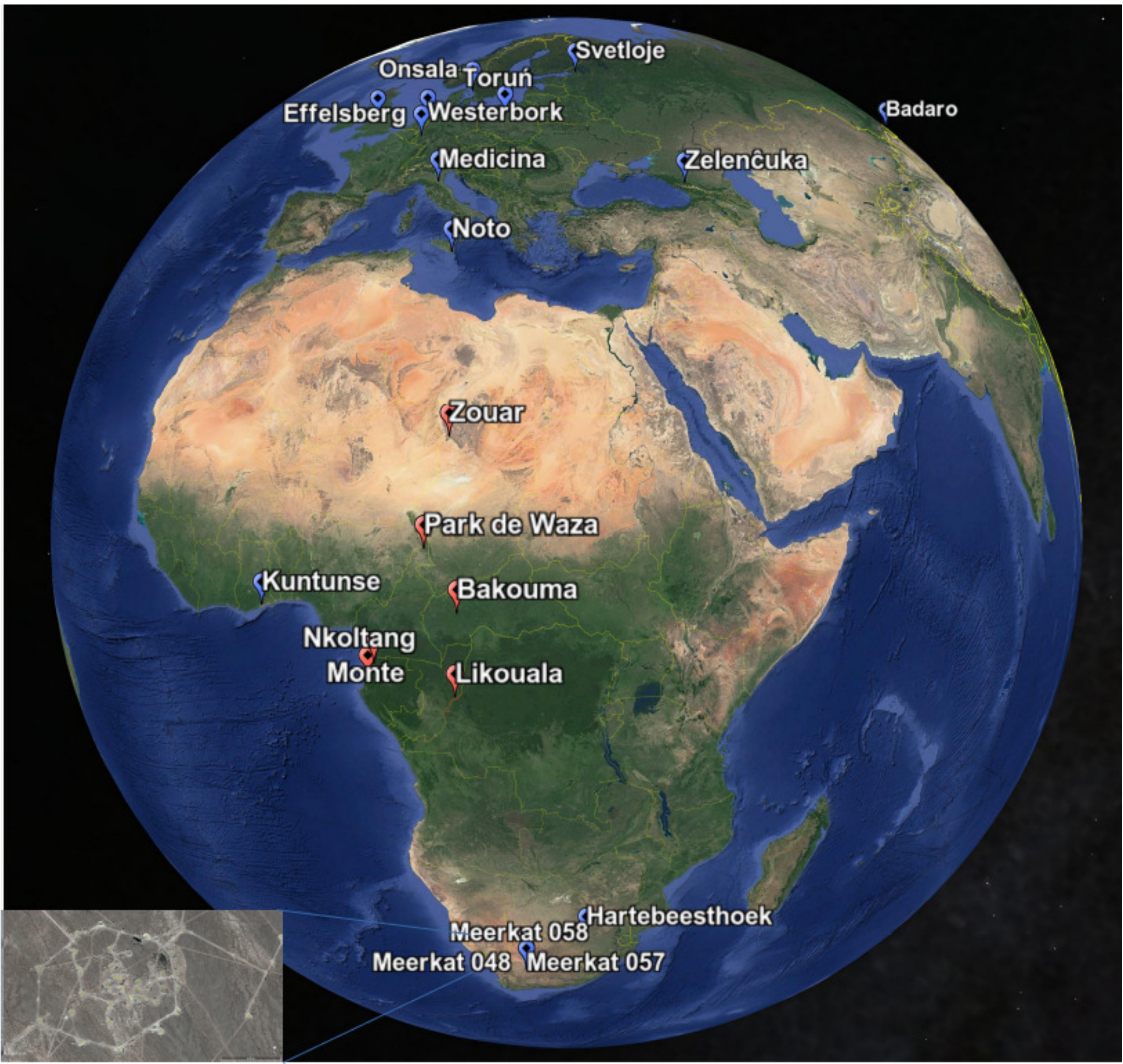}
    \caption{Blue points: Locations of the  Kuntunse antenna in Ghana, the MeerKAT stations in South Africa and the EVN. Red points: 
    Locations of abandoned old telecommunication satellite facilities in the ECCAS area and/or  possible sites to build new radio telescopes.}
  \label{fig:coffee}
\end{figure}

In this section, we discuss the performances in $uv$-coverage density of the combined Kuntunse antenna in Ghana with the MeerKAT telescope in South Africa, correlated to the full EVN. The full EVN consists of 
12 stations across the globe i.e.  Badary, Effelsberg, Hartebeesthoek, Jodrell Bank, Medicina, Noto, Onsala, Shanghai, Svetloe, Torun, Westerbork and Zelenchukskaya. 
Figure~\ref{fig:coffee} shows an African map where the blue points are the Kuntunse antenna, the MeerKAT in South Africa and the EVN. There are some stations of the EVN that do not appear on the map (e.g. Shanghai), these are stations that are in the other side of the globe. The 64 antennas of the MeerKAT do not appear all on the map, this is because the antennas are very close to each other making it difficult to visualise in a bigger scale. 
The points in red  are locations of abandoned old telecommunications satellites in the ECCAS countries or locations suitable to build new radio antennas.

\begin{figure}
  \centering
    \includegraphics[width=0.5\textwidth]{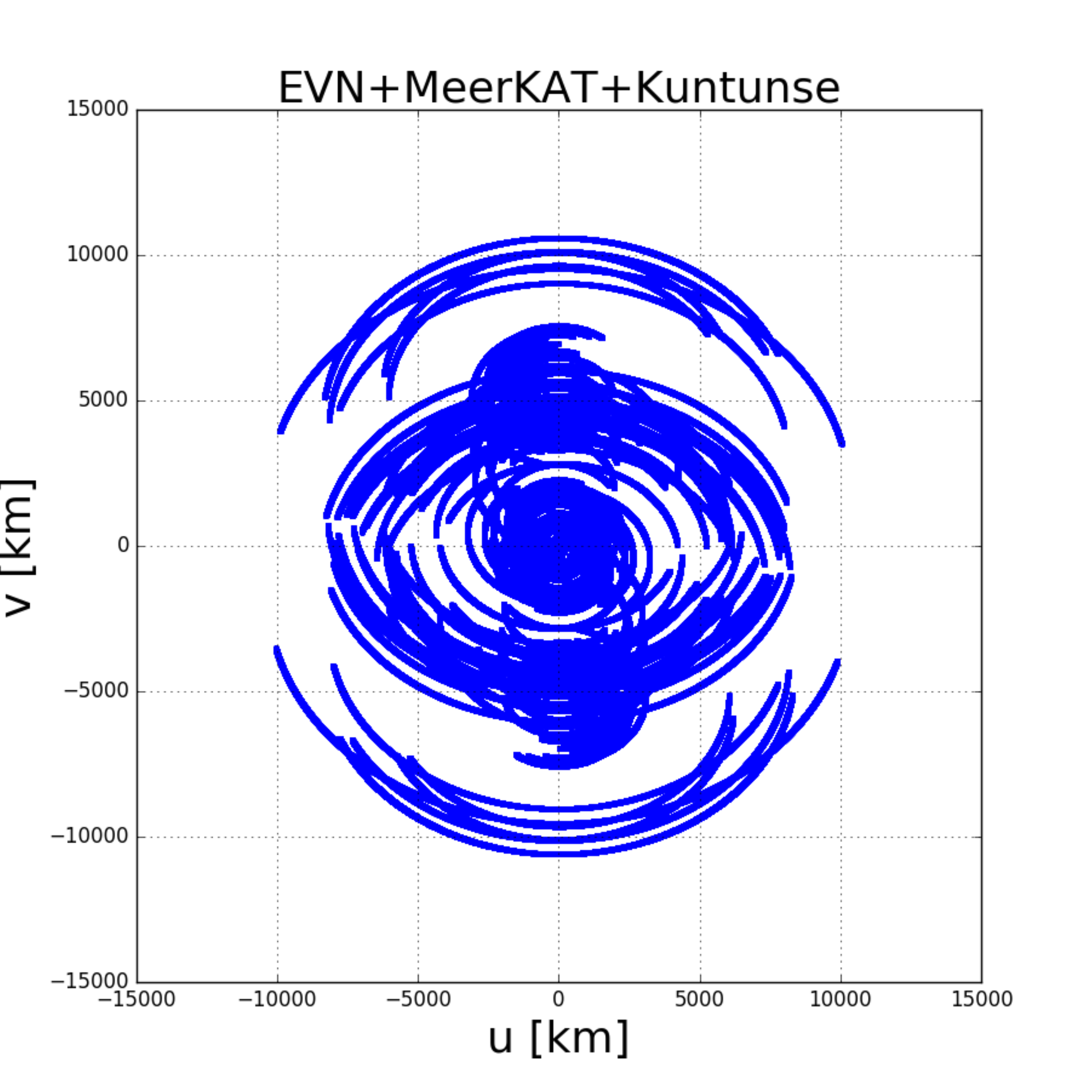}
    \caption{The EVN combined with the AVN (MeerKAT and Kuntunse antenna in Ghana).}
  \label{fig:coffee1}
\end{figure}

Figure~\ref{fig:coffee1} shows 
the $uv$-coverage of Kuntunse combined with the MeerKAT and both correlated to the EVN. This $uv$-coverage is obtained by simulation at a frequency of 16 GHz,  during a total period of 10 hours with 1 s integration time and 256 MHz bandwidth divided into channels of 100 kHz each.
This $uv$-coverage is tracking a source at a declination of +45 deg. We made sure that all the antennas are able to see the source at this position.
It is clearly seen from this $uv$-coverage that there are missing samples in the middle area i.e. the area in between the core and the outer core of the $uv$-coverage. The samples from the core are from shorter baselines; 
these shorter baselines are the internal baselines of the MeerKAT telescope and the EVN. The samples at the outer core are from the longer baselines;  these relate the AVN antennas in the upper hemisphere to MeerKAT and Hartebeesthoek in the southern hemisphere.
There are few medium-length baselines coming from the correlation between the antennas in the upper hemisphere or southern hemisphere to the Kuntunse antenna in Ghana. 
To fill these missing samples, we need more medium-length baselines. These medium-length baselines can only be obtained if some of the antennas are placed around the
equatorial line in Africa.
Most of the countries in Africa around the equator are the French-speaking countries or the ECCAS countries. Using simulations we show in this work that if one were to build radio telescopes and/or convert old abandoned telecommunication satellite antennas to radio telescopes in the ECCAS countries no matter where these antennas are to be located in each of these countries, this should improve the current AVN $uv$-coverage significantly. 

\section{Existing abandoned telecommunication facilities}
The potential for converting obsolete large antennas for radio astronomy has been recognized World-wide. 
The opportunity exists for African countries to re-use these initially expensive assets that have become redundant with the advent of the optical fiber.
Such thrusts in infrastructure revamp could lead to the development of fertile paths for local research in radio astronomy and space science as well as 
promote science development on the continent, at a relatively low cost. The SKA Africa partner countries that currently host such redundant large antennas are South Africa (3 antennas), Ghana, Kenya, Madagascar and Zambia. Similar large antennas have been located in Algeria (2), Benin, Cameroon (2), Congo Peoples Republic, Egypt (2),
Ethiopia, Malawi, Morocco, Niger, Nigeria (3), Senegal, Tunisia, Uganda, Gabon and Zimbabwe. The antenna in Mozambique was dismantled and probably also the one in Togo. 
Gabon is a country in the ECCAS  community currently hosting such idling ground station.
We use it here as a benchmark for describing the typical fate of other existing dishes in the ECCAS region. The satellite telecommunication antenna in Gabon was commissioned in  early 1973 as a node of access into the global network of Intelsat Satellite Earth Stations. The latter was part of a thrust initiated in the 1960's when communication via satellites orbiting the Earth was introduced to carry voice, data and TV signals, to supplement undersea cables. The radio bands allocated for these satellite communications were $5.925 - 6.425$ GHz for uplink and $3.700 - 4.200$  GHz for downlink. 
These are within the frequency range commonly referred to as  ``C-band''.  Initially a Standard A antenna had to be at least $30$ meters in diameter, and the antennas built in Africa from 1970 to 1985 are this size~\citep{Gaylard}. Exactly as the one in Gabon visible in Figure~\ref{TIG} of Appendix A.   
\section{Simulations and results}
Two simulations are  performed using  antennas as showed in Figure~\ref{fig:coffee}. Firstly, we simulate the $uv$-coverage using only the 6 antennas of the ECCAS countries i.e. antennas with position marked in red in Figure~\ref{fig:coffee}  by evaluating the performance 
of this single interferometer if this were to be used to observe the sky individually. Secondly, we correlate the 6 ECCAS antennas with the AVN and the EVN. The latter demonstrates the scientific advantages of adding these 6 antennas to the current VLBI network. 

\subsection{Performance assessment of the $uv$-coverage of the ECCAS antennas}
\begin{figure}[]
    \centering
    \begin{minipage}{.5\textwidth}
        \centering
        \includegraphics[width=1.\linewidth, height=0.3\textheight]{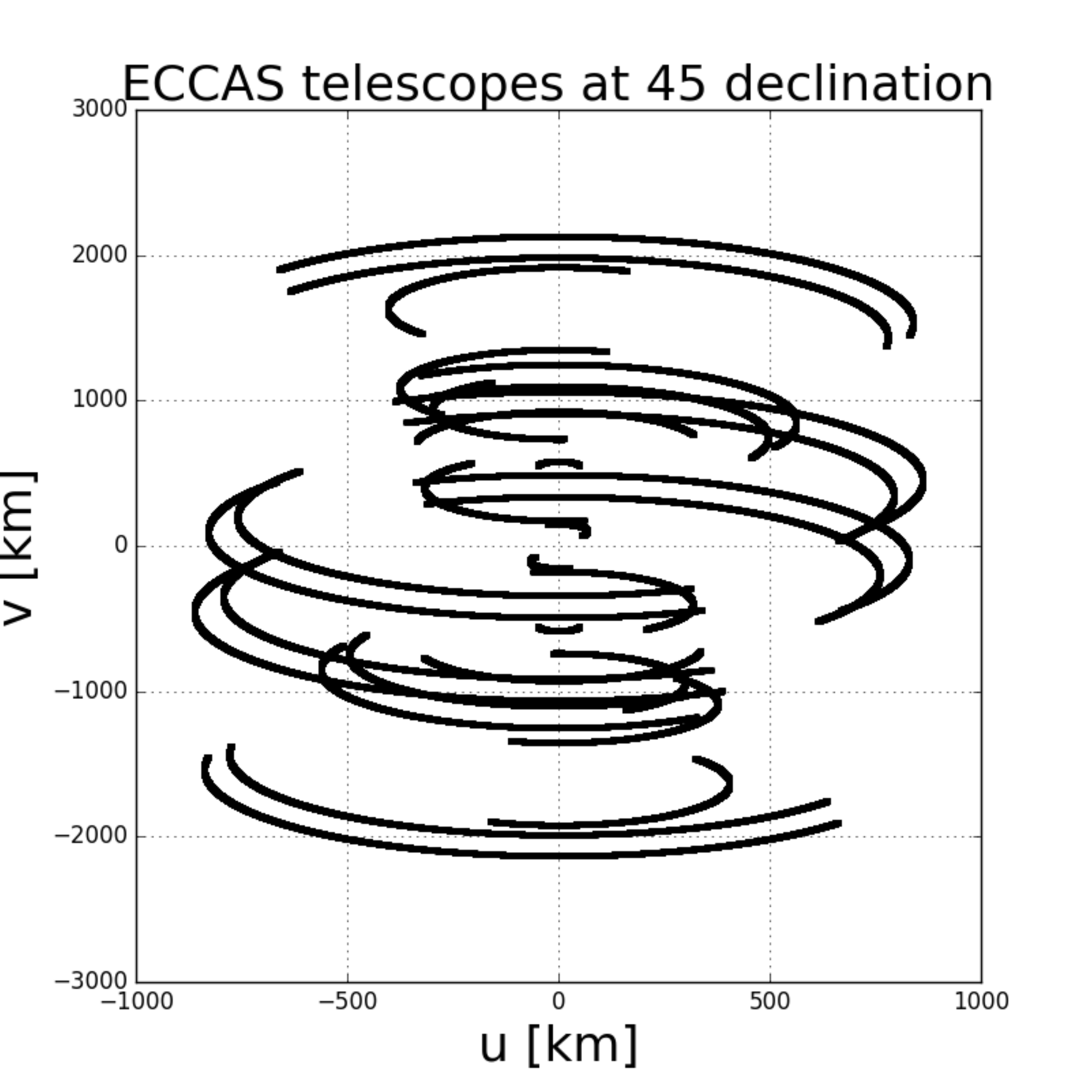}
    \end{minipage}%
    \begin{minipage}{0.5\textwidth}
        \centering
        \includegraphics[width=1.\linewidth, height=0.3\textheight]{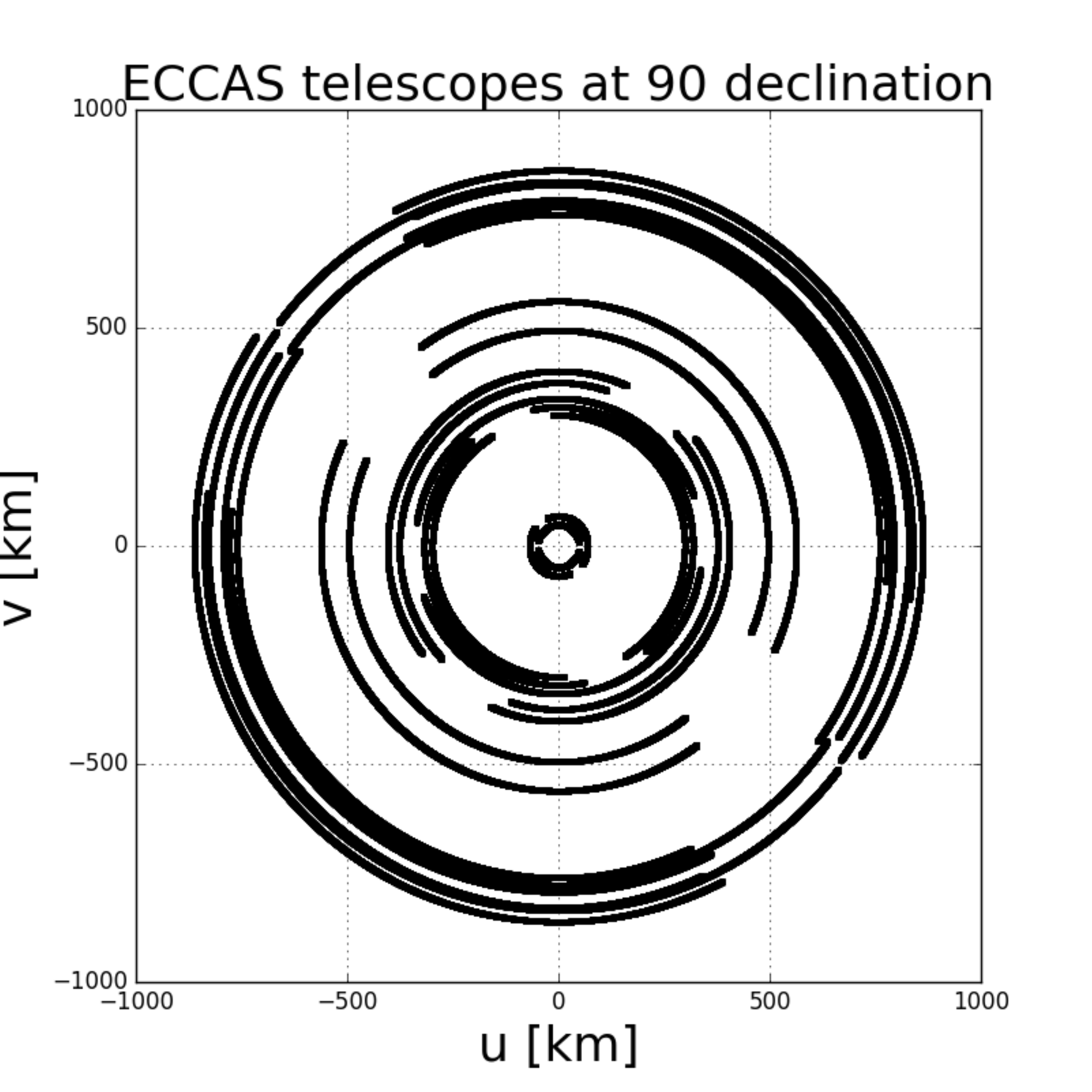}
    \end{minipage}
    \caption{ECCAS antennas $uv$-coverage at 1.4 GHz at two declinations ($45^{o}$ and $90^{o}$), 10 hr observation and 256 MHz total bandwidth showing a lot of holes or gaps.}
    \label{fig:cemmac}
\end{figure}

Figure~\ref{fig:cemmac} shows the $uv$-coverage of the  6 antennas in the ECCAS countries at two declinations. This $uv$-coverage was simulated during 10 hr total time with 1 s integration time and total bandwidth of 256 MHz divided into channels  of width 100 kHz. 
The
positions of the antennas are shown in Figure \ref{fig:coffee}, red points. As expected, the $uv$-coverage is very
poor as the 6 antennas are spread over a large distance. Each of these antennas can function as a single dish radio
telescope and can do high level science e.g., observing pulsars, masers, hot gas from the
Milky Way or distant galaxies. They will thus significantly broaden the science area of the
local Universities. The radio telescope in Ghana has already demonstrated its ability to do
real science by observing methanol masers and pulsars. These telescopes can also be used
to train local students and also serve as outreach facilities to motivate students and local
residents to be interested in science. Finally, the telescopes can join the geodesy, geodynamics
and astrometry research with the VLBI network, thereby broadening its scientific
research capabilities.


\subsection{Filled $uv$-coverage for  ECCAS + MeerKAT + EVN + Kuntunse}
\begin{figure}
  \centering
    \includegraphics[width=0.5\textwidth]{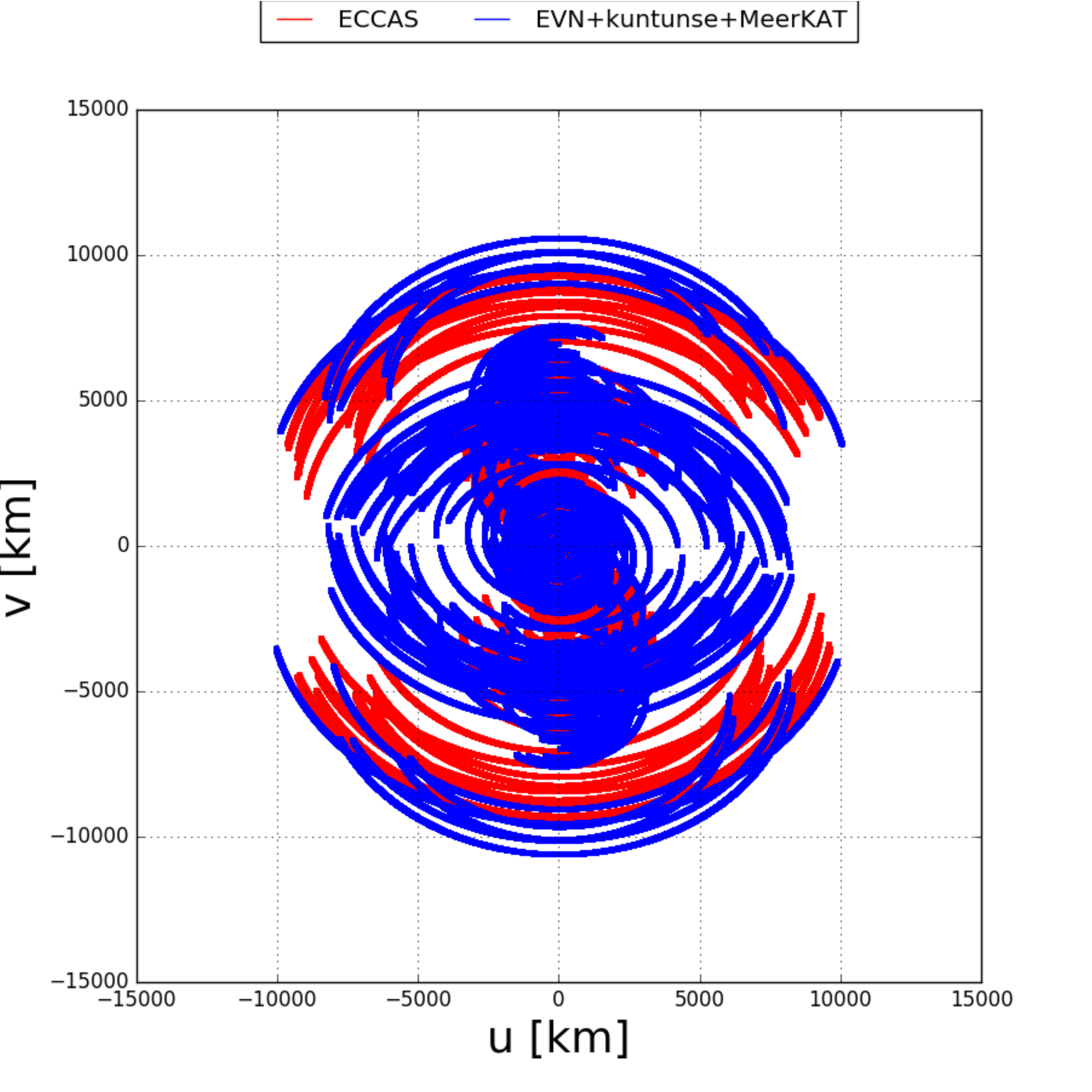}
    \caption{Performance of the global VLBI $uv$-coverage. The ECCAS antennas are correlated with the Kuntunse antenna, the MeerKAT and the EVN. 
    The $uv$-coverage is now well filled because of the extra medium-length baselines that relate the EVN and MeerKAt to the ECCAS antennas.}
  \label{fig:TIG1}
\end{figure}

This section describes the main results of this paper. The simulation presented in section~\ref{sect1} is repeated. This time the MeerKAT telescope, EVN and Kuntunse are correlated to the ECCAS antennas. The full antennas used in the simulation are shown in Figure~\ref{fig:coffee}. 
Figure~\ref{fig:TIG1} shows the principal result achieved in this work. The blue points in Figure~\ref{fig:TIG1} are the data from the EVN, MeerKAT and Kuntunse antenna, while the red points are data coming from  the ECCAS antennas and their correlation to the EVN, MeerKAT and Kuntunse. 
We note that while the 6 ECCAS antennas on their own give poor $uv$-coverage as presented in Figure~\ref{fig:cemmac}, they significantly improve the current $uv$-coverage of the full VLBI network.

\section{Conclusion}
We conclude that the $uv$-coverage for a full VLBI observation will improve if few antennas were to be added in the ECCAS region. The resulting Fourier transform of the $uv$-coverage compactness will only lead to low confusion noise limit which is suitable for high signal-to-noise or dynamic range images requirement.  Building or converting old abandoned satellites telecommunication facilities in the ECCAS region
is a guarantee that the science results from these antennas will expand the universities international visibility.
As part of the VLBI, the scientific community of the ECCAS region will  fully be prepared in a strong scientific involvement with the SKA.

\vspace{1cm}

\subsection*{Acknowledgement}
This work is based upon research supported by the South African
Research Chairs Initiative of the Department of Science and Technology
and National Research Foundation.
The MeerKAT telescope is operated by the South African Radio Astronomy Observatory, which is a facility of the National Research
Foundation, an agency of the Department of Science and Innovation.
We thank our colleagues Dr Aard Keimpema at the Joint Institute for VLBI ERIC and Dr Griffin Foster at the University of Oxford for their insights and comments on early drafts of this work. MA is funded by Rhodes University. 
P.O is funded by the University of South Africa (UNISA).

\section*{Appendix A: Decommissioned 32 meter large Satellite Earth Station antennas in Gabon}
Today, while remaining the property of the state, the idling dish above is part of the infrastructure 
currently leased to a private operator in the mobile telephony sector. It is located in an area known as 
Nkoltang, in the Northern part of Libreville. The dish is adjacent to the telemetry ground station used by 
the French CNES to perform follow up of ARIANE launch from French Guyana. 
If refurbished, idling satellite Earth station antennas such as the one in Gabon will operate 
from existing facilities relatively close to cities in relatively populated neighborhood.
The issue of relatively high level of pre-existing Radio Frequency Interference (RFI) will 
therefore need to be dealt with appropriately.
Standard procedures currently in place in  Kuntunse  easily serve as a template. Kuntunse, 
in Ghana is currently the site hosting the first ever idling ground station in Africa that has been successfully 
refurbished to become an operational radio-telescope. In the absence of dedicated protocols, the dominant approach 
there consists in performing standard flagging during the data analysis with standard packages. Such approach has the 
advantage to provide further opportunities for building capacity in standard  procedures using standard resources often 
very well supported by the comnunity. Coupled with the advent of MOOC\footnote{Massive open online course}, the initial 
demand on expert human capital as trainers is therefore minimized.

\begin{figure}
  \centering
    \includegraphics[width=0.6\textwidth]{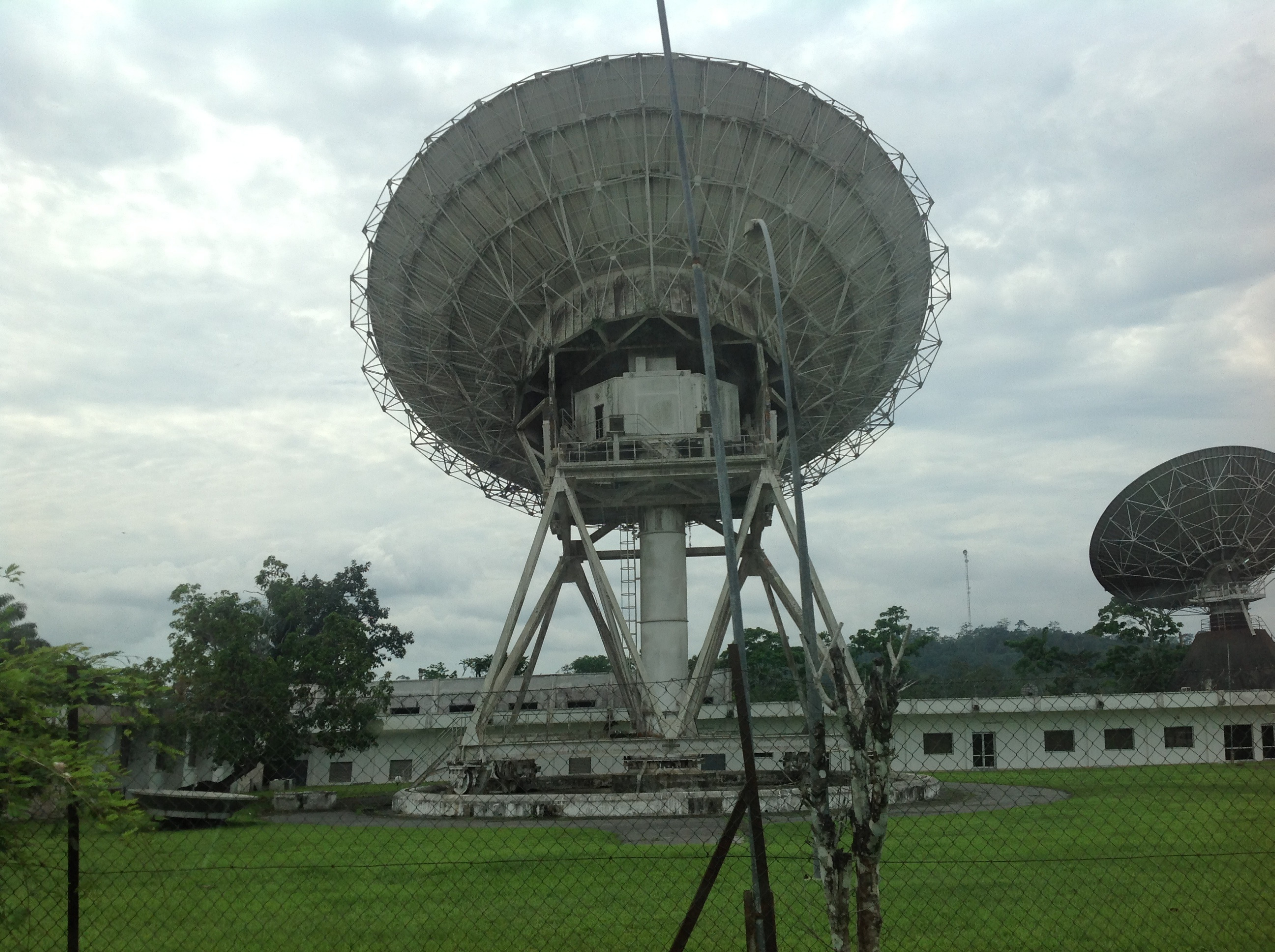}
    \caption{Decommissioned 32 meters large Satellite Earth Station antennas in Gabon.}
  \label{TIG}
\end{figure}

\end{document}